\documentclass[twocolumn,aps,prb,10pt]{revtex4-1}
\pdfoutput=1
\usepackage{graphicx}
\usepackage{color}
\usepackage{tikz}
\usepackage{pgfplots} 
\usepackage{amssymb}
\pgfplotsset{tick scale binop=\times}

\begin{document}
\title{Bias Cancellation in One-Determinant Fixed-Node Diffusion Monte Carlo: Insights from Fermionic Occupation Numbers}
\author{Mat\'u\v{s} Dubeck\'{y}}\email{matus.dubecky@osu.cz}
\affiliation{Department of Physics, Faculty of Science, University of Ostrava, 30. dubna 22, 701 03 Ostrava, Czech Republic}
\affiliation{ATRI, Faculty of Materials Science and Technology, Slovak University of Technology, Paul\'inska 16, 917 24 Trnava, Slovakia}

\date{\today}

\begin{abstract}
{
Accuracy of the fixed-node diffusion Monte Carlo (FN-DMC) depends on the node location of the best available trial state $\Psi_T$.
The practical FN-DMC approaches available for large systems rely on compact yet effective $\Psi_T$s containing explicitly correlated single Slater determinant (SD). 
However, SD nodes may be better suited to one system than to another, which may possibly lead to inaccurate FN-DMC energy differences. It remains a challenge, how to estimate inequivalency or appropriateness of SDs.
Here we use the differences of a measure based on Euclidean distance between the natural orbital occupation number (NOON) vector of the Slater determinant (SD) from the exact solution in the NOON vector space, that  
can be viewed as a measure of SD inequivalency and a measure of the expected degree of nondynamic-correlation-related bias in FN-DMC energy differences.
This is explored on a set of small noncovalent complexes and covalent bond breaking of Si$_2$ vs. N$_2$. It turns out that NOON-based measures well reflect the magnitude and sign of the bias present in the data available, thus providing new insights to the nature of bias cancellation in SD FN-DMC energy differences.
}
\end{abstract}
\maketitle

\section{Motivation}
Fixed-node diffusion Monte Carlo (FN-DMC) is a projector many-body electronic structure method\cite{Anderson1975,Reynolds1982,Umrigar1993}, promising for its accuracy, massive parallelism, low-order CPU cost scaling and direct treatment of extended models\cite{Kolorenc2011rev,Austin2012rev}. For a given Hamiltonian $H$, FN-DMC projects out the ground-state $\Psi$ that has non-zero overlap with the antisymmetric trial state  $\Psi_T$, in imaginary \mbox{time $\tau$:}
\begin{equation}
\Psi= \lim_{\tau\rightarrow\infty} \exp(-\tau H)\Psi_T.
\end{equation}
In real space, FN-DMC captures symmetric correlations exactly, and the accuracy of Fermi states is limited by the location of the supplied approximate node, i.e., a subset of electron positions $\mathbf{R}$ where $\Psi_T(\mathbf{R})=0$. The total FN-DMC energy is an upper bound to the exact energy\cite{Moskowitz1982} (which would result if the node would be exact) and the related bias, FN bias, scales quadratically with the nodal displacement error\cite{Mitas1991}. In principle, $\Psi_T$ can be systematically improved so that the related FN bias becomes negligible\cite{Morales2012,Clay2015}, however, with increasing system size, such an approach eventually becomes unreliable. 

The practical FN-DMC approaches available for large systems\cite{Ambrosetti2014,Benali2014} therefore rely on compact yet effective $\Psi_T$s like Slater-Jastrow (SJ) ans\"atze\cite{Jastrow1955}, containing explicitly correlated single Slater determinant (SD)\cite{Ceperley1977} (or a small number of them). Such an approach makes FN-DMC somewhat empirical, because the nodes of SJ $\Psi_T$s  are not converged and related FN bias is hard to control. Thus, the method performance must be mapped a priori for a representative class of systems considered\cite{Grossman2002,Petruzielo2012}. In general, nevertheless, FN-DMC using SJ wave functions is predictive and reaches acceptable accuracy in a number of important systems where it has no competitors, e.g., transition metal oxides at high pressure\cite{Kolorenc2008}, magnetic states in solids\cite{Caffarel2014,Wagner2014,Wagner2015,Reboredo2014,Zheng2015} or large noncovalent systems\cite{Tkatchenko2012,Ambrosetti2014,Benali2014,Mostaani2015,Hamdani2015} (for more examples, see reviews \citenum{luchow2000,Foulkes2001rev,Lester2009rev,
Bajdich2009rev,needs2010,Kolorenc2011rev,Luchow2011rev,Austin2012rev,Morales2014rev,Wagner2014rev,Brown2014rev,Dubecky2014rev}).

In small noncovalent systems, surprisingly accurate single-point FN-DMC interaction energies of the order of 0.1 kcal/mol  vs. benchmark CCSD(T) results extrapolated to complete basis set (CBS)\cite{Gurtubay2007,Santra2008,Ma2009,Korth2011,Gillan2012,Dubecky2013,Dubecky2014,Benali2014,Hamdani2014} have been reported. Such a high level of accuracy obtained with far-from-accurate SJ trial wave functions, was attributed to a high degree of FN bias cancellation operative in systems dominated by ``sufficiently'' weak interactions\cite{Mella2003,Diedrich2005,Dubecky2013}, taking place in low-density\cite{Rasch2012,Kulahlioglu2014,Rasch2014} intermolecular regions with a predominant $s$ character where the exchange contributions are negligible\cite{Dubecky2016rev}.  
Bias cancellation limits of such an approach are however unknown, that somewhat hampers widespread usage of FN-DMC as a benchmark method. Deviations of the order of 0.3 kcal/mol (relative errors exceeding 7\%), unacceptable at the benchmark level (relative errors should not exceed say 2-3\%), of unknown nature found in complexes with multiply bound hydrogen bond acceptors like HCN dimer or formaldehyde (FD) dimer\cite{Dubecky2014} start to delineate its limits and motivate our further elaborations in this field.

In multireference systems, it is known that SDs do not suffice for FN-DMC\cite{Rasch2012,Kulahlioglu2014}, and, SJ $\Psi_T$s should contain additional dominant configurations so that the FN bias well cancels out\cite{Zimmerman2009,Bouabca2009,Berner2010,Valsson2010,Zen2015b}. On the other hand, in closed-shell systems, the effect of additional determinants is expected to be negligible. Nevertheless, this largely depends on the target accuracy, and, for HCN and FD with a target high-standard accuracy (0.1 kcal/mol), we suspect, that the observed FN-DMC discrepancies stem from the inequivalent treatment of nondynamic correlation effects\cite{Sinanoglu1964,Matito2016} in $\Psi_T$s of the molecular clusters vs. their constituents.
Since SD FN-DMC strictly neglects nondynamic correlations, contrary to single-reference coupled cluster (CC) that employs the low-rank substitutions, the related bias becomes non-negligible in energy-differences even in ``single-reference'' cases where the contributions of additional configurations to the states of interacting vs. non-interacting systems are sufficiently different.
An extreme example would be to use a SD FN-DMC in cases with non-negligible near-degeneracy like atomization with spin uncoupling (e.g., N$_2\rightarrow$N+N). It is known that such cases require large number of determinants (that vary for incident systems/states\cite{Powell2016,Dubecky2011} because of different convergence rate\cite{Clay2015,Deible2015}), to achieve converged FN-DMC energy differences\cite{Bajdich2010b,Petruzielo2012}, simply because the individual (ground) states (of N$_2$ and N in our example) are of different nature owing to their spin multiplicity and near-degeneracy. 

Clearly, SD nodes from the best available theory may be better suited to one system than to another, which would cause non-systematic bias compensation and possibly inaccurate FN-DMC energy differences. For a certain accuracy level, it remains a challenge, how to estimate inequivalency or appropriateness of SDs used in FN-DMC.


In this work, we demonstrate that nondynamic-correlation-related bias cancellation in SD FN-DMC can be qualitatively understood in terms of the differences of a measure based on an Euclidean distance between the exact solution (or a good approximation to it) and the Slater point (related SD occupation) in the natural orbital (NO) occupation number (ON) vector space\cite{Chakraborty2014,Tennie2016}. 

The NOON-space distance between the state and its best-matching Slater point is usually related to the amount of nondynamic/static/strong\cite{Sinanoglu1964,Bartlett2007,Hollett2011,Matito2016} correlation in the considered state\cite{Chakraborty2014,Chakraborty2015}. Nevertheless, the same distance can also be viewed as a property of a given SD, namely its ``distance from exactness''. Differences of such distances for a set of related states (e.g., for interacting vs. non-interacting system), obtained within an equivalent theory, may be, at least qualitatively, viewed as a measure of SD inequivalency and consequently as a measure of the expected degree of bias related to such SDs, if used as reference or trial states (assuming that inequivalencies propagate to energy differences). 

This connection  motivates our qualitative inequivalency analysis of Slater parts of SJ wave functions used in FN-DMC. For sets of states where accurate NOONs can be obtained from post-Hartree-Fock (post-HF) theories, it enables a priori estimation of the expected degree of nondynamic-correlation-related bias cancellation in SD SJ FN-DMC energy differences. 
Below, we explore such a possibility for a set of small noncovalent complexes including benzene dimer. The defined measures provide new insights to FN bias cancellation, that are extensively discussed along with the possible limitations and improvements. In addition, an example of covalent bond breaking (Si$_2$ vs. N$_2$) is also considered.

\section{Definitions}\label{secdef}

Below, we focus on energy differences (noncovalent interaction energies and covalent bond breaking energies) defined as
\begin{equation}
\Delta E = E^C - E^S,
\end{equation}
where $E^S$ and $E^C$ are the total energies of the supersystem $S$ (e.g. molecular cluster or molecule with covalent bond) and its non-interacting counterpart $C$  
\begin{figure}[h!]
 \centering
\begin{tikzpicture}
\draw[thick] (2,2) circle (0.3);
\draw[thick] (4,2) circle (0.3);
\draw[dashed,thick] (2.4,2) -- (3.6,2);
\node[align=left] at (6.5,2) {$S$, interacting};
\draw[thick] (2,0.7) circle (0.3);                                                                                                                                                                                \draw[thick] (4,0.7) circle (0.3);
\node[align=left] at (6.5,0.7) {C, non-interacting};
\end{tikzpicture}
\end{figure}
(e.g. molecules/atoms constituting a system before covalent/non-covalent bond dissociation $S$, where the fragments may be sufficiently separated so that interaction is negligible or treated independently) with geometries same as in $S$ (i.e., the deformation energy is not taken into account),
respectively.

We assume that the difference of FN-DMC interaction energy ($\Delta E^\mathrm{FN}$) from the exact result ($\Delta E$),
\begin{equation}
 \epsilon=\Delta E^\mathrm{FN}-\Delta E,
\end{equation}
can be expressed as a sum of contributions,
\begin{equation}
 \epsilon = \epsilon_\mathrm{stat} + \epsilon_\mathrm{sam} + \epsilon_\mathrm{FN} + \epsilon_\mathrm{ECP} + \epsilon_\mathrm{tstep} + \epsilon_\mathrm{other},
\end{equation}
where $\epsilon_\mathrm{stat}$ is a stochastic bias, $\epsilon_\mathrm{sam}$ is a sampling error related to non-ergodicity of electron dynamics due to $\Psi_T$s, $\epsilon_\mathrm{FN}$ is a FN bias, $\epsilon_\mathrm{ECP}$ relates to the quality of ECP constructions and treatment of ECPs in FN-DMC, $\epsilon_\mathrm{tstep}$ is a finite time-step bias, and, other yet unknown possible biases are denoted as $\epsilon_\mathrm{other}$. 

The NOONs, i.e. the eigenvalues of the one-particle reduced density matrix $\rho(\mathbf{r},\mathbf{r}')$ \cite{Lowdin1955}, can be obtained from the diagonalization of a discrete representation $\alpha$ corresponding to a state expanded in an orthonormal basis of one-particle spatial orbitals $\{\phi_i\}$,
\begin{equation}
 \rho(\mathbf{r}',\mathbf{r})=\sum_{ij}\alpha_{ij}\phi_i(\mathbf{r})\phi_j^*(\mathbf{r'}),
\end{equation}
leading to
\begin{equation}
 \rho(\mathbf{r}',\mathbf{r})=\sum_{i}\lambda_{i}\psi_i(\mathbf{r})\psi_j^*(\mathbf{r'}).
\end{equation}
Here $\{\psi_i\}$ is a set of NOs and $\{\lambda_i\}$ is a set of the corresponding NOONs normalized to the number of considered electrons $N$, $\sum_{i=1}^K\lambda_i=N$, that reflect the Pauli\cite{Pauli1925} and generalized Pauli principle\cite{Klyachko2006,Schilling2013}, and, account for the spin-degeneracy: $0\leq\lambda_i\leq 2$. In the following, we use the $\vec{\lambda}$ vectors ordered in ascending way, i.e., $\lambda_1\geq \lambda_2\geq\dots\lambda_{K-1}\geq\lambda_K$, where $K$ is the total number of one-particle basis functions.

For a given state, we define an Euclidean distance\cite{Chakraborty2014,Tennie2016},
\begin{equation}
 d'(\vec\lambda,\vec\lambda^\mathrm{SD})=\|\vec{\lambda}-\vec{\lambda}^\mathrm{SD}\|  = \sqrt{\sum_i (\lambda_i-\lambda^\mathrm{SD}_i)^2},
\end{equation}
between its $\vec{\lambda}$-vector (for practical purposes, obtained from the qualitatively correct correlated size-consistent post-HF computation), and, the Slater point, $\vec{\lambda}^\mathrm{SD}$, here corresponding to the occupation of SD part of SJ $\Psi_T$ used in \mbox{FN-DMC}. For instance, for a closed-shell ground-state, \mbox{$\vec{\lambda}^\mathrm{SD}=\{2,2,\dots,2,0,\dots,0\}$}. 
As such, $d'$ measures nondynamic correlation content of a given state, or, equivalently, distance of SD from the exact solution in NOON vector space. The determinantal part of $\Psi_T$ fixes the node locus, and, convincing evidence shows that FN-DMC recovers dynamic correlations exactly within the node constraint and the nodes are directly responsible for the remaining missing nondynamic correlations (see, e.g., refs.\citenum{Berner2010,Rasch2012,Kulahlioglu2014}). We therefore expect that systems where SD $\Psi_T$ causes large FN bias, $d'$ will be also large in magnitude, and vice versa.
Note, that  one-particle orbitals of the SD part can be adjusted without affecting $\vec{\lambda}^\mathrm{SD}$. Therefore, relevant to the present discussion are only the orbitals that minimize FN-DMC energy, at least approximately. We assume that intact DFT orbitals suffice for this purpose, and, only nondynamic correlations that cannot be recovered within SD FN-DMC by any means\cite{DubJur} contribute to FN bias and its differences. 


We rescale $d'$ so that it becomes size-intensive,
\begin{equation}
 d=\frac{d'}{\sqrt{N}},
\end{equation} 
where, from now on, $N$ is the number of explicitly correlated electrons in post-HF treatment.

We define a differential measure of nondynamic correlations, or SD inequivalency, relevant to energy differences,
\begin{equation}\label{dd}
 \delta=d^C-d^S,
\end{equation}
that is based on scaled distances $d^S$ and $d^C$ obtained for the supersystem $S$, and its non-intereacting counterpart $C$, respectively. We expect larger $|\epsilon_\mathrm{FN}|$ in systems where $|\delta|$ is larger and vice versa. One may compare $\delta$s to the differences of another near-degeneracy measure known from CC theory, the so-called  $T_1$ diagnostic\cite{Lee1989}, $\Delta T_1=T_1^C-T_1^S$.

Finally, we define a tentative empirical measure valid for molecular dimers in equilibrium,
\begin{equation}\label{eqr}
 \kappa=\delta B_1 B_2,
\end{equation}
where $B_i$ is a maximum bond multiplicity of a contact atom corresponding to the constituent $i$. The purpose of $B_i$s is to qualitatively reflect the expected rise of FN bias with higher bond multiplicity\cite{Rasch2014}. In linear HCN dimer, for instance, we set $B_1=3$ and $B_2=1$ since the hydrogen bond (contact) consists of triply-bound nitrogen and hydrogen.

\section{Methods and Models} 

Below, we demonstrate the utility of the introduced quantities, $d$, $\delta$, and $\kappa$, computed at the frozen-core CCSD/aug-VDZ level. We assume that CCSD theory provides accurate NOONs in single-reference cases but also in $s/p$ atoms. The convergence of $d$ and $\delta$ vs. the basis set saturation was explicitly checked: the \mbox{aug-VDZ} and aug-VTZ data were indistinguishable so that we assume that NOONs are basis-set converged at the aug-VDZ level. 

We consider a set of eleven small single-reference closed-shell complexes, four larger complexes containing benzene (Bz), and two covalent main group dimers (for a full list, see Tab.~\ref{tabres}), where the statistically converged SD FN-DMC results employing Jastrows with electron-nucleus, electron-electrons and electron-electron-nucleus terms are available\cite{Dubecky2013,Dubecky2014,Petruzielo2012}. The FN-DMC energy differences are compared vs. the reference CCSD(T)/CBS\cite{Rezac2015,Takatani2010} data, and we assume that for our purposes, it is reasonable to assume that $\Delta E \approx \Delta E^\mathrm{CCSD(T)/CBS}$. 

Note that our CCSD results are based on HF reference determinant but they are expected to relate well to our FN-DMC results using DFT trial states, because the orbital relaxation effects in CC are taken into account via $T_1$ operator (see, e.g., ref.~\citenum{Noga2009}). Additional tests with Brueckner orbitals show negligible differences of our NO-based measures with respect to CC computations employing canonical HF orbitals.

\section{Small Noncovalent Complexes}
%
\begin{figure}[t!]
\begin{flushright}
\centering	
\begin{tikzpicture}[scale=0.9] \begin{axis}[ width=0.52\textwidth, height=0.2\textwidth, xtick pos=left,
ytick pos=left, ybar, enlargelimits=0.1, x tick label style={rotate=90}, every y tick scale label/.style={at={(.1,.85)}},
ylabel={$d^S$, $d^C$},xtick=\empty,bar width=5]
\addplot[fill=yellow,bar shift = -3.0] coordinates {
(1,  3.18E-02)
(2,  3.66E-02)
(3,  4.21E-02)
(4,  4.51E-02)
(5,  4.86E-02)
(6,  2.77E-02)
(7,  3.17E-02)
(8,  2.85E-02)
(9,  2.28E-02)
(10, 4.81E-02)
(11, 3.80E-02)
(13,4.00E-2)
(14,3.95E-2)
(15,4.20E-2)
(16,4.19E-2)
};
\addplot[fill=cyan,bar shift = 2.0] coordinates {
(1,  3.17E-02)
(2,  3.65E-02)
(3,  4.21E-02)
(4,  4.52E-02)
(5,  4.87E-02)
(6,  2.77E-02)
(7,  3.16E-02)
(8,  2.86E-02)
(9,  2.30E-02)
(10, 4.84E-02)
(11, 3.87E-02)
(13,4.00E-2)
(14,3.95E-2)
(15,4.19E-2)
(16,4.19E-2)
};
\end{axis} \end{tikzpicture}
\begin{tikzpicture}[scale=0.9] \begin{axis}[ width=0.52\textwidth, height=0.2\textwidth, xtick pos=left,
ytick pos=left, ybar, enlargelimits=0.1, x tick label style={rotate=90}, every y tick scale label/.style={at={(.1,.85)}},
ylabel={$T_1^S$, $T_1^C$},xtick=\empty,bar width=5]
\addplot[fill=yellow,bar shift = -3.0] coordinates {
(1,   6.90E-03)
(2,   8.13E-03)
(3,   1.01E-02)
(4,   1.10E-02)
(5,   1.20E-02)
(6,   1.02E-02)
(7,   9.93E-03)
(8,   1.24E-02)
(9,   1.24E-02)
(10,  1.41E-02)
(11,  1.79E-02)
(13,8.92E-3)
(14,1.02E-2)
(15,9.30E-3)
(16,9.43E-3)
};
\addplot[fill=cyan,bar shift = 2.0] coordinates {
(1,   6.76E-03)
(2,   8.21E-03)
(3,   1.02E-02)
(4,   1.11E-02)
(5,   1.21E-02)
(6,   1.00E-02)
(7,   9.69E-03)
(8,   1.23E-02)
(9,   1.25E-02)
(10,  1.40E-02)
(11,  1.67E-02)
(13,8.98E-3)
(14,1.02E-2)
(15,9.48E-3)
(16,9.49E-3)
};
\end{axis} \end{tikzpicture}
\begin{tikzpicture}[scale=0.9] \begin{axis}[ width=0.52\textwidth, height=0.2\textwidth, xtick pos=left,
ytick pos=left, ybar, enlargelimits=0.1, x tick label style={rotate=90}, every y tick scale label/.style={at={(.1,.45)}},
ylabel={$\Delta T_1$},xtick=\empty]
\addplot[fill=blue,bar shift = 0.0] coordinates {
(1,   -1.43E-04)
(6,   -1.82E-04)
(7,   -2.33E-04)
(8,   -5.86E-05)
(10,  -8.35E-05)
(11,  -1.18E-03)
(14,-4.64E-5)
};
\addplot[fill=red,bar shift = 0.0] coordinates {
(2,   8.27E-05)
(3,   2.16E-05)
(4,   1.23E-04)
(5,   1.09E-04)
(9,   1.25E-04)
(13, 5.34E-5)
(15, 1.81E-4)
(16, 5.83E-5)
};
\end{axis} \end{tikzpicture}
\begin{tikzpicture}[scale=0.9] \begin{axis}[ width=0.52\textwidth, height=0.2\textwidth, xtick pos=left,
ytick pos=left, ybar, enlargelimits=0.1, x tick label style={rotate=90}, every y tick scale label/.style={at={(.1,.85)}},
ylabel={-$\epsilon$},xtick=\empty,]
\addplot[fill=red,bar shift = 0.0] coordinates {
(4,    0.06)
(5,    0.10)
(8,    0.08)
(9,    0.10)
(10,   0.38)
(11,   0.33)
(14, 0.24)
(16, 0.17)
};
\addplot[fill=blue,bar shift = 0.0] coordinates {
(1,   -0.00)
(2,   -0.14)
(3,   -0.01)
(6,   -0.11)
(7,   -0.05)
 (13,-0.15)
 (15,-0.32)
};
\end{axis} \end{tikzpicture}
\begin{tikzpicture}[scale=0.9] \begin{axis}[ width=0.52\textwidth, height=0.2\textwidth, xtick pos=left,
ytick pos=left, ybar, enlargelimits=0.1, x tick label style={rotate=90}, every y tick scale label/.style={at={(.1,.85)}},
ylabel={$\delta$},xtick=\empty,]
\addplot[fill=red,bar shift = 0.0] coordinates {
(4,    9.30E-05)
(5,    9.03E-05)
(8,    1.29E-04)
(9,    1.95E-04)
(10,   2.65E-04)
(11,   6.51E-04)
(13, 0)
(14, 6.05E-5)
(15,-9.94E-5)
(16, 4.26E-6)
};
\addplot[fill=blue,bar shift = 0.0] coordinates {
(1,    -9.15E-05)
(2,    -5.39E-05)
(3,    -1.04E-05)
(6,    -3.35E-05)
(7,    -6.23E-05)
(13,-3.71E-5)
(15,-9.94E-5)
};
\end{axis} \end{tikzpicture}
\begin{tikzpicture}[scale=0.9] \begin{axis}[ width=0.52\textwidth, height=0.2\textwidth, xtick pos=left,
ytick pos=left, ybar, enlargelimits=0.1, x tick label style={rotate=90},  every y tick scale label/.style={at={(.1,.85)}},
ylabel={$\kappa$}, symbolic x coords={CH$_4$ dimer,Ethene-Ar,Ethene dimer,Ethene-ethyne,Ethyne dimer,CH$_4$-FH,NH$_3$ dimer,H$_2$O dimer,FH dimer,HCN dimer,FD dimer,~,Bz-CH$_4$,Bz-H$_2$O,Bz$_2^\mathrm{P}$,Bz$_2^\mathrm{T}$}, xtick=data]
\addplot[fill=red,bar shift = 0.0] coordinates {
(CH$_4$ dimer, 0)
(Ethene-Ar,    0)
(Ethene dimer, 0)
(Ethene-ethyne, 1.86E-04)
(Ethyne dimer,  2.71E-04)
(CH$_4$-FH,    0)
(NH$_3$ dimer, 0)
(H$_2$O dimer,  1.29E-04)
(FH dimer,      1.95E-04)
(HCN dimer,     7.96E-04)
(FD dimer,      2.60E-03)
(Bz-CH$_4$,          0)
(Bz-H$_2$O,           1.21E-4)
(Bz$_2^\mathrm{P}$,  0)
(Bz$_2^\mathrm{T}$,   1.28E-5)
};
\addplot[fill=blue,bar shift = 0.0] coordinates {
(CH$_4$ dimer,  -9.15E-05)
(Ethene-Ar,     -1.08E-04)
(Ethene dimer,  -4.16E-05)
(CH$_4$-FH,     -3.35E-05)
(NH$_3$ dimer,  -6.23E-05)
(Bz-CH$_4$,          -7.43E-5)
(Bz$_2^\mathrm{P}$,  -3.98E-4)
};

\end{axis} \end{tikzpicture}
\end{flushright}
\caption{Comparison of FN-DMC and CCSD/aug-VDZ results for a set of noncovalent complexes: deviations from the reference $\epsilon$ (kcal/mol), $T_1$ diagnostics for dimolecular complexes $S$ (supersystem, yellow) and non-interacting constituents $C$ (cyan) and their differences $\Delta T_1$, rescaled NOON-space distances of the correlated post-HF NOONs from the Slater point ($d$), their differences ($\delta$), and, the corresponding bond-multiplicity-rescaled differences $\kappa$. For definitions, see Sec.\ref{secdef}.}
\label{figres}
\end{figure}
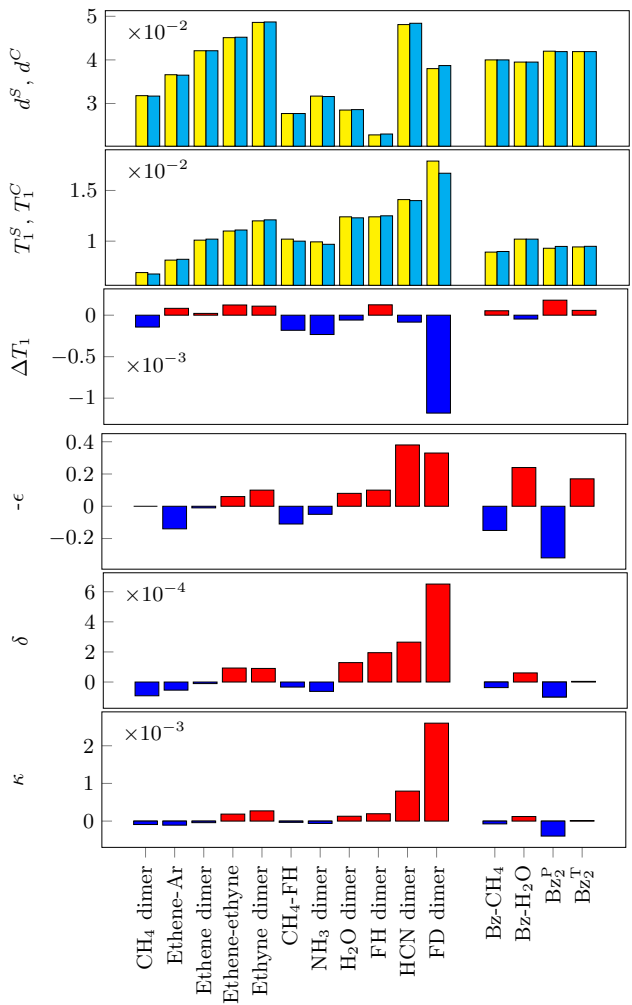
%

The results for eleven small noncovalent complexes are reported in Fig.~\ref{figres} (left set) and Tab.~\ref{tabres} (topmost part).
In these systems, $d$s are typically smaller for the trivially saturated closed-shell complexes and higher for complexes containing $\pi$ bonds, and change only mildly between $S$ and $C$ (Fig.~1, $d^S$ vs. $d^C$). The largest values were found for the ethyne dimer and HCN dimer, both containing a maximum (triple) bond multiplicity. This is rather promising, since, for instance, the $T_1$ diagnostic does not reveal such intuitively acceptable trends here. E.g., $T_1$s for FH are larger than for some $\pi$-complexes.

We now turn to the quantities relevant to energy differences. The measure $\delta$ provides the key insights of this work.
It appears that in cases where FN-DMC works well, i.e. where $|\epsilon|$ is small (Fig.~\ref{figres}, $\delta$ vs. $\epsilon$), $|\delta|$ is also small. In systems with larger discrepancies, $|\delta|$ is larger, and it is most prominent for HCN and FD where $|\epsilon|$ is largest. $\delta$ thus well informs on inequivalency of SDs and suggests likely source of bias (see below). For comparison, $\Delta T_1$ does not show any clear trends with respect to the FN-DMC data available.

The results reported to date for small single-reference\cite{Rezac2015} closed-shell noncovalent complexes\cite{Dubecky2013,Dubecky2014} lead us to believe, that the discrepancies $\epsilon$ observed in HCN and FD dimers\cite{Dubecky2014} relate to $\epsilon_\mathrm{other}$, or, more likely, to $\epsilon_\mathrm{FN}$ caused by non-equivalency of nondynamic correlations omitted in SD $\Psi_T$s and consequent less systematic FN bias cancellation in energy differences, in contrast to the complexes where $|\delta|$ is small. Other sources of bias in $\epsilon$ can likely be ruled out: $\epsilon_\mathrm{stat}$ by sufficient convergence of error bars ($\leq0.07$~kcal/mol), $\epsilon_\mathrm{sam}$ is diminished by the use of correlation-adjusted (DFT) orbitals\cite{Per2012,Clay2015} and diffuse functions\cite{Dubecky2013}, $\epsilon_\mathrm{ECP}$ is negligible by the quality of ECP constructions\cite{Rezac2015} and the use of three-body Jastrow factors\cite{Dubecky2014} in $\Psi_T$s, and, the time-step bias does not appear to be the root cause\cite{Dubecky2014}. 

At the current stage, it looks that for HCN and FD, SD FN-DMC with DFT orbitals reaches its limits. Indeed, one of the ways to achieve a higher accuracy is a systematic improvement of $\Psi_T$s by adding more determinants, but, this approach is of little interest for large systems. Alternatively, one can explore more systems and hope for universal trends as a function of interaction strength and/or bond multiplicity and develop some sort of empirical correction based on $\delta$.

It is interesting to observe that the sign of $\delta$ alters in the same way as $\epsilon$ in {\it all} the considered cases.  
Positive $\delta$ consistent with overbinding shows up in cases where the state of the system $C$ has more near-degeneracy than the state corresponding to $S$, a case of HCN and FD.  Here SD FN-DMC recovers more correlation energy in system $S$ than in $C$, making the production energy difference overestimated. On the other hand, if $\delta$ is larger for $S$, the energy difference is expected to be underestimated by the analogous reasoning. This clarifies why SD FN-DMC sometimes under- and sometimes overestimates the energy differences.

Although $\delta$ is a promising quantity, it is a bit disturbing that it is sometimes similar for complexes where FN-DMC behaves in a qualitative different way. For instance, $\delta$ in trivially saturated FH dimer, where SD FN-DMC leads to benchmark results\cite{Rezac2015}, is only slightly smaller than $\delta$ in HCN dimer, where the method exceedingly overestimates $\Delta E$.
On one hand, this may indicate that the total energies used for evaluation of $\Delta E^\mathrm{FN}$ in HCN dimer are not converged and can be somehow further improved (one can imagine, that, e.g., the Jastrow-term optimization stuck in local minimum), or, more likely, as mentioned above, $\epsilon$ suggests that $\epsilon_\mathrm{FN}$ is larger in HCN dimer than in FH dimer and $\delta$ does not provide sufficient resolution for these two cases. It would be desirable to develop a quantity that would discern between such cases and allow for assessment of FN-DMC data without resorting to reference $\Delta E$s. 

We thus propose to use another measure, based on $\delta$, reflecting also the fact that, in general, FN errors depend on electron density\cite{Rasch2014}. Although this requires a deeper study, as an example, we define here an empirical measure $\kappa$ that progressively depends on a maximum bond multiplicity of terminal atoms constituing noncovalent contact/s (Eq.~\ref{eqr}). This quantity can be used to sort the complexes according to the empirically chosen critical value (here $\kappa_\mathrm{crit}=5\times10^{-4}$) into two categories:  safe $(\kappa\ll\kappa_\mathrm{crit})$, and, possibly problematic $(\kappa\gtrsim\kappa_\mathrm{crit})$. Visually, on the scale of Fig.~\ref{figres} (left set), the two classes emerge: ``small'' bars, and, ``appreciable'' bars (dimers of HCN, FD). $\kappa$ thus presumably indicates, that the quality of FN-DMC results in HCN and FD cannot be guaranteed to such an accuracy like for the remaining complexes. We note, that analogs of $\kappa$ would be desired for blind assessment of FN-DMC energy differences, and, indeed, more rigorous and general definition/s (valid for potential surface scans or clusters with more contacts) should explicitly depend on electron density (e.g., at bond critical points) instead of $B$s. 

\section{Larger Noncovalent Complexes}
We now test the introduced measures in larger, Bz-containing complexes, Bz-H$_2$O, \mbox{Bz-CH$_4$}, parallel-displaced Bz$_2$ (Bz$_2^\mathrm{P}$), and, T-shaped Bz$_2$ (Bz$_2^\mathrm{T}$). According to $\delta$ and $\kappa$, SD FN-DMC should underestimate $\Delta E$ in Bz-CH$_4$ and Bz$_2^\mathrm{P}$, and, overestimate in Bz-H$_2$O and Bz$_2^\mathrm{T}$ (Fig.~\ref{figres}, right set, and, Tab.~\ref{tabres}). We find that this is fully in agreement with the best available $\epsilon$s. According to $\kappa$ ($B_\mathrm{Bz}=2$), accurate energy differences are expected for Bz-CH$_4$,  Bz-H$_2$O, and, Bz$_2^\mathrm{T}$, and the biases observed in these complexes are most probably of statistical nature. On the other hand, more pronounced $\kappa$ in Bz$_2^\mathrm{P}$ indicates that its $\Delta E^\mathrm{FN}$  may be somewhat underestimated by SD FN-DMC, which is in agreement with the data available\cite{Azadi2015,Gasperich2016}. Interestingly, for the two variants of Bz$_2$, $\kappa$ shows that Bz$_2^\mathrm{P}$ has a more pronounced differential multireference character that makes the FN bias compensation of SD FN-DMC in this complex less efficient. $\kappa$ thus suggests why SJ FN-DMC in Bz$_2^\mathrm{T}$  produces results in a good agreement with the reference\cite{Dubecky2013}, while the same is not so easy for Bz$_2^\mathrm{P}$\cite{Azadi2015,Gasperich2016}.

Note, however, that although $\delta$ or $\kappa$ is largest for Bz$_2^\mathrm{P}$ from the four Bz-containing complexes considered, so that the largest error is expected for this complex, the absolute values of $\delta$/$\kappa$ are much smaller that those for FD or even HCN. More specifically, $\delta$ and $\kappa$ for Bz$_2^\mathrm{P}$ is comparable to the same in complexes where FN-DMC works well. Therefore, the current results do not conclusively answer the question, whether Bz$_2^\mathrm{P}$ can or cannot be well described within SD FN-DMC framework, but leave room for future attempts.

\section{Covalent Bond Breaking}

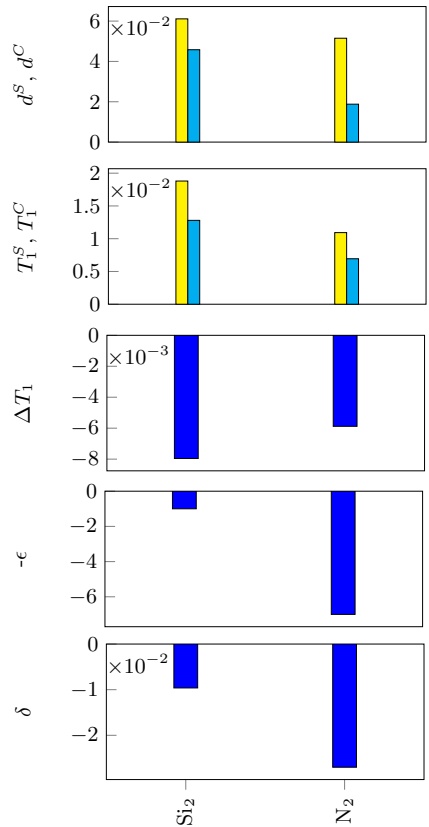
\begin{figure}[ht!]
\begin{flushright}
\centering	
\begin{tikzpicture}[scale=0.9] \begin{axis}[ width=0.35\textwidth, height=0.2\textwidth, xtick pos=left,
ytick pos=left, ybar, enlarge x limits=0.5, x tick label style={rotate=90}, every y tick scale label/.style={at={(.1,.85)}},
ylabel={$d^S$, $d^C$},xtick=\empty,bar width=5,ymin=0]
\addplot[fill=yellow,bar shift = -2.5] coordinates {
(1,  6.11e-2)
(2,  5.15e-2)
};
\addplot[fill=cyan,bar shift = 2.5] coordinates {
(1,  4.58E-02)
(2,  1.88E-02)
};
\end{axis} \end{tikzpicture}
\begin{tikzpicture}[scale=0.9] \begin{axis}[ width=0.35\textwidth, height=0.2\textwidth, xtick pos=left,
ytick pos=left, ybar, enlarge x limits=0.5, x tick label style={rotate=90}, every y tick scale label/.style={at={(.1,.85)}},
ylabel={$T_1^S$, $T_1^C$},xtick=\empty,bar width=5,ymin=0]
\addplot[fill=yellow,bar shift = -2.5] coordinates {
(1,   1.88E-02)
(2,   1.093E-02)
};
\addplot[fill=cyan,bar shift = 2.5] coordinates {
(1,   1.28E-02)
(2,   6.94E-3)
};
\end{axis} \end{tikzpicture}
\begin{tikzpicture}[scale=0.9] \begin{axis}[ width=0.35\textwidth, height=0.2\textwidth, xtick pos=left,
ytick pos=left, ybar, enlarge x limits=0.5, x tick label style={rotate=90}, every y tick scale label/.style={at={(.1,.85)}},
ylabel={$\Delta T_1$},xtick=\empty,ymax=0]
\addplot[fill=blue,bar shift = 0.0] coordinates {
(1,   -7.96E-03)
(2,  -5.88E-03)
};
\end{axis} \end{tikzpicture}
\begin{tikzpicture}[scale=0.9] \begin{axis}[ width=0.35\textwidth, height=0.2\textwidth, xtick pos=left,
ytick pos=left, ybar, enlarge x limits=0.5, x tick label style={rotate=90}, every y tick scale label/.style={at={(.1,.85)}},
ylabel={-$\epsilon$},xtick=\empty,ymax=0.0]
\addplot[fill=blue,bar shift = 0.0] coordinates {
(1,   -1.0)
(2,   -7.0)
};

\end{axis} \end{tikzpicture}
\begin{tikzpicture}[scale=0.9] \begin{axis}[ width=0.35\textwidth, height=0.2\textwidth, xtick pos=left,
ytick pos=left, ybar, enlarge x limits=0.5, x tick label style={rotate=90},  every y tick scale label/.style={at={(.1,.85)}},
ylabel={$\delta$}, symbolic x coords={Si$_2$,N$_2$}, xtick=data,ymax=0]
\addplot[fill=blue,bar shift = 0.0] coordinates {
(Si$_2$, -9.62e-3)
(N$_2$,   -2.7e-2)
};
\end{axis} \end{tikzpicture}
\end{flushright}
\caption{Comparison of FN-DMC and CCSD/aug-VDZ results for Si$_2$ and N$_2$. For description, see caption of Fig.~\ref{figres}.}
\label{figres2}
\end{figure}

Finally, we explore the performance of $\delta$ applied to atomization of N$_2$ vs. Si$_2$, where SD FN-DMC shows vastly different deviations from the reference atomization energies\cite{Petruzielo2012}. 
While in Si$_2$ FN-DMC underbinds by about 1 kcal/mol, it does even more so in N$_2$ where $\epsilon$ achieves as much as 7 kcal/mol (Fig.~\ref{figres2}, Tab.~\ref{tabres}). Although $d$s are comparable to the above-mentioned noncovalent systems (Tab.~\ref{tabres}), $\delta$s are significantly more pronounced here, that is consistent  with much larger bias observed in energy differences (including the correct sign). In addition, $\delta$ for N$_2$ is almost three times larger than the value for  Si$_2$, indicating significantly less efficient FN bias cancellation in this case, a trend fully consistent with the available energy differences (Fig.~\ref{figres2}, $\epsilon$ vs. $\delta$). Altogether, for atomization of the two considered molecules,  consistency of $\delta$ with the available FN-DMC data presumably indicates that $\delta$-based measures may find practical use also in systems with a pronounced strong correlation character, including covalent bond-breaking scenarios. Regarding the $\kappa$, its definition in terms of $B$s requires more care, and we postpone its evaluations to the follow-up work. 

\begin{table*}[h!]
\caption{Results: FN-DMC interaction energy deviations from the CCSD(T)/CBS reference $\epsilon$ (kcal/mol), the number of correlated electrons $N$, and, contact atom bond multiplicities $B_i$.  The results corresponding to the CCSD/aug-VDZ level of calculations for the given supersystem $S$ (or complex) and its non-interacting constituents $C$ include $T_1$ diagnostics and their differences $\Delta T_1$, rescaled NOON distances from the Slater point ($d$), their differences ($\delta$), and, bond-multiplicity-rescaled differences $\kappa$. For definitions, see Sec.\ref{secdef}.}
\begin{tabular}{l|ccc|cc|ccccccccc}
\hline
\hline
                 &    & &       & \multicolumn{2}{c}{FN-DMC}& \multicolumn{7}{|c}{CCSD/aug-VDZ}     \\
Complex          &$N$& $B_1$& $B_2$& $\epsilon$ & Ref.  & $T_1^S$ & $T_1^C$ & $\Delta T_1$& $d^S$ & $d^C$ & $\delta$ & $\kappa$ \\
\hline
CH$_4$ dimer	 &16& 1 & 1 &  0.00& \citenum{Rezac2015}     & 6.90E-3&6.76E-3&-1.43E-4       &3.18E-2&3.17E-2&-9.15E-5 & -9.15E-5\\
Ethene-Ar	 &20& 2 & 1 &  0.14& \citenum{Rezac2015}     & 8.13E-3&8.21E-3& 8.27E-5       &3.66E-2&3.65E-2&-5.39E-5 & -1.08E-4\\
Ethene dimer	 &24& 2 & 2 &  0.01& \citenum{Dubecky2013}   & 1.01E-2&1.02E-2& 2.16E-5       &4.21E-2&4.21E-2&-1.04E-5 & -4.16E-5\\
Ethene-ethyne	 &22& 2 & 1 & -0.06& \citenum{Dubecky2013}   & 1.10E-2&1.11E-2& 1.23E-4       &4.51E-2&4.52E-2& 9.30E-5 &  1.86E-4\\
Ethyne dimer	 &20& 3 & 1 & -0.10& \citenum{Rezac2015}     & 1.20E-2&1.21E-2& 1.09E-4       &4.86E-2&4.87E-2& 9.03E-5 &  2.71E-4\\
CH$_4$-FH	 &16& 1 & 1 &  0.11& \citenum{Rezac2015}     & 1.02E-2&1.00E-2&-1.82E-4       &2.77E-2&2.77E-2&-3.35E-5 & -3.35E-5\\
NH$_3$ dimer	 &16& 1 & 1 &  0.05& \citenum{Dubecky2013}   & 9.93E-3&9.69E-3&-2.33E-4       &3.17E-2&3.16E-2&-6.23E-5 & -6.23E-5\\
H$_2$O dimer	 &16& 1 & 1 & -0.08& \citenum{Dubecky2013}   & 1.24E-2&1.23E-2&-5.86E-5       &2.85E-2&2.86E-2& 1.29E-4 &  1.29E-4\\
FH dimer	 &16& 1 & 1 & -0.10& \citenum{Rezac2015}     & 1.24E-2&1.25E-2& 1.25E-4       &2.28E-2&2.30E-2& 1.95E-4 &  1.95E-4\\
HCN dimer	 &20& 3 & 1 & -0.38& \citenum{Rezac2015}     & 1.41E-2&1.40E-2&-8.35E-5       &4.81E-2&4.84E-2& 2.65E-4 &  7.96E-4\\
FD dimer	 &24& 2 & 2 & -0.33& \citenum{Rezac2015}     & 1.79E-2&1.67E-2&-1.18E-3       &3.80E-2&3.87E-2& 6.51E-4 &  2.60E-3\\
\hline
Bz-CH$_4$	 &38& 2 & 1 &  0.15& \citenum{Dubecky2013}   & 8.92E-3&8.98E-3& 5.34E-5       &4.00E-2&4.00E-2&-3.71E-5 & -7.43E-5\\
Bz-H$_2$O	 &38& 2 & 1 & -0.24& \citenum{Dubecky2013}   & 1.02E-2&1.02E-2&-4.64E-5       &3.95E-2&3.95E-2& 6.05E-5 &  1.21E-4\\
Bz$_2^\mathrm{P}$&60& 2 & 2 &  0.32& \citenum{Azadi2015}     & 9.30E-3&9.48E-3& 1.81E-4       &4.20E-2&4.19E-2&-9.94E-5 & -3.98E-4\\
Bz$_2^\mathrm{T}$&60& 2 & 2 & -0.17& \citenum{Dubecky2013}   & 9.43E-3&9.49E-3& 5.83E-5       &4.19E-2&4.19E-2& 4.26E-6 &  1.28E-5\\
\hline
N$_2$	         &10& 1 & 1 &  7.0 & \citenum{Petruzielo2012}&1.28E-2&6.94E-3& -5.88E-3       &4.58E-2&1.88E-2&-2.70E-2 & -2.70E-2\\
Si$_2$	         & 8& 1 & 1 &  1.0 & \citenum{Petruzielo2012}&1.88E-2&1.09E-2& -7.96E-3       &6.11E-2&5.15E-2&-9.62E-3 & -9.62E-3\\
\hline\hline
\end{tabular}
\label{tabres}
\end{table*}

\section{Summary}
NOON-based  measures, proportional to the differential amount of nondynamic correlations omitted in SD $\Psi_T$s, were used to assess the expected degree and sign of bias cancellation in SD FN-DMC energy differences. The use of NOONs guarantees generality and universality across various size-consistent methods (e.g. CCSD, CAS-SCF, full CI). The new measures, here tested on reference CCSD data, revealed rapid convergence with a basis set, thus enabling their use in reasonably large complexes (e.g. benzene dimer), proved their utility in application to small single-reference closed-shell noncovalent complexes, and, they suggested the nature of biases in dimers of HCN and FD, benzene-containing complexes, and, atomization of N$_2$ vs. Si$_2$. More work is required to assess their performance in larger number of small complexes, larger complexes, and in combination with other theories (e.g., higher-order CC, CAS-SCF, full CI or DMRG). 

~

\begin{acknowledgments}
The author is grateful to Lubos Mitas,  Petr Jure\v{c}ka, and, Daniel Plencner, for fruitful discussions and useful comments on the manuscript. This work was financially supported by University of Ostrava (UO, IRP201558) and VEGA (project Nos. 1/0770/13 and 2/0130/15). The calculations were performed at local facility of UO (purchased from EU funds, project No. CZ.1.05/2.1.00/19.0388), Metacentrum CESNET (LM2015042) and CERIT (LM2015085), and, TACC under XSEDE allocation (provided by Lubos Mitas). This research also used a Director's Discretionary allocation at the Argonne Leadership Computing Facility, which is a DOE Office of Science User Facility supported under Contract DE-AC02-06CH11357. Kind assistance by Dr. Anouar Benali is gratefully acknowledged.
\end{acknowledgments}


\end{document}